\documentclass[11pt]{amsart}
\usepackage{amsmath}

\begin{document}

\title[Measuring the shape of the universe]{Measuring the
shape of the universe}
\author[Neil J. Cornish \& Jeffrey R. Weeks]{Neil
J. Cornish$^{\diamondsuit}$
and Jeffrey R. Weeks$^{\dag}$}

\thanks{$\diamondsuit\, $Department of Applied Mathematics and Theoretical
Physics, University of Cambridge, Silver Street, Cambridge CB3 9EW, UK}
\thanks{\dag\ 88 State St, Canton NY 13617, USA}

\maketitle

\section*{Introduction}

Since the dawn of civilization, humanity has grappled with the big
questions of existence and creation. Modern cosmology seeks to answer
some of these questions using a combination of mathematics and
measurement. The questions people hope to answer include ``how did the
universe begin?''; ``how will the universe end?''; ``is space finite
or infinite?''. After a century of remarkable progress, cosmologists
may be on the verge of answering at least one of these
questions -- is space finite?  Using some simple geometry and
a small NASA satellite set for launch in the year 2000,
the authors and their colleagues hope to measure the size and shape
of space. This article explains the mathematics behind the
measurement, and the cosmology behind the observations.

Before setting out, let us first describe the broad picture we have in
mind. Our theoretical framework is provided by Einstein's theory of
general relativity and the hot big-bang model of cosmology. General
relativity describes the universe in terms of geometry, not just of
space, but of space and time. Einstein's equation relates the
curvature of this space-time geometry to the matter contained in the
universe.

A common misconception is that
the curvature of space is all one needs to answer the question
``is space finite or infinite?''. While it is true that spaces of
positive curvature are necessarily finite, spaces of negative or zero
curvature may be either finite or infinite.
In order to answer questions about the global
geometry of the universe we need to know both its curvature and
topology. Einstein's equation tells us nothing about the topology of
spacetime since it is a local equation relating the spacetime curvature
at a point to the matter density there.
To study the topology of the universe we need to measure
how space is connected. In doing so we will not only discover whether
space is finite, but also gain insight into physics beyond
general relativity.

The outline of our paper is as follows. We begin with an introduction
to big bang cosmology, followed by a review of some basic
concepts in geometry and topology. With these preliminaries out of
the way, we go on to describe the plan to measure the size and
shape of the universe using detailed observations of the
afterglow from the big bang.

\section*{Big bang cosmology}

The big bang model provides a spectacularly successful description of our
universe. The edifice is supported by three main observational
pillars: (1) the uniform expansion of the universe; (2) the abundances
of the light elements; (3) the highly uniform background of microwave
radiation.

The primary pillar was discovered by Edwin Hubble in the early 1920s.
By comparing the spectral lines in starlight from nearby and
distant galaxies, Hubble noticed that the vast majority of
distant galaxies have their spectra shifted to the red, or long
wavelength, part of the electromagnetic spectrum. Moreover, the
redshift was seen to be larger for more distant galaxies, and to occur
uniformly in all directions. A simple explanation for this observation
is that the space between the galaxies is expanding isotropically.
By the principle of mediocrity -- {\it i.e.} we do
not live at a special point in space -- isotropic expansion
about each point implies homogenous expansion. Such a homogeneous
and isotropic expansion can be characterized by an overall
scale factor $a(t)$ that depends only on time. As the universe expands,
the wavelength $\lambda$ of freely propagating light is stretched
so that
\begin{equation}\label{lam}
\lambda(t_0) = \lambda(t) { a(t_0) \over a(t)} \, ,
\end{equation}
where $t_0$ denotes the present day and $t$
denotes the time when the light was emitted.
Astronomers define the {\it redshift} $z$ as the fractional change
in the wavelength:
\begin{equation}\label{red}
z = {\lambda(t_0) - \lambda(t) \over \lambda(t)}  \, .
\end{equation}
Since we expect atoms to
behave the same way in the past, we can use atomic spectra measured on
Earth to fix $\lambda(t)$.
Using equation (\ref{lam}) we can relate the redshift to the size of
the universe:
\begin{equation}\label{az}
a = { a_0  \over (1+z)} \, .
\end{equation}
We have adopted the standard shorthand $a_0 = a(t_0)$
for denoting quantities measured today,
and $a = a(t)$ for denoting quantities measured at a generic time $t$.
By measuring the redshift of an object we can infer how big the
universe was when the light was emitted. The relative size of
the universe provides us with a natural notion of time in
cosmology. Astronomers like to use redshift $z$ as a measure of
time ($z=0$ today, $z=\infty$ at the big
bang) since, unlike the time $t$, the redshift is a measurable quantity.

A photon's energy $E$ varies inversely with its wavelength $\lambda$.
A gas of photons at temperature $T$ contains photons with energies in
a narrow band centered at an energy $E$ that is proportional to
the temperature. Thus $T \sim E \sim \lambda^{-1}$, and
the temperature of a photon gas evolves as
\begin{equation}\label{temp}
{ T \over T_0} = {E \over E_0}= {\lambda_0 \over \lambda}=
{a_0 \over a}=1+z \, .
\end{equation}
This equation tells us that the universe should have been much hotter
in the past than it is today. It should also have been much denser.
If no particles are created or destroyed, the density of ordinary
matter is inversely proportional to the occupied volume, so it
scales as $\rho_m \sim a^{-3}$. If no photons are created or
destroyed, the number of photons per unit volume also scales as
$a^{-3}$. However, the energy of each photon is decreasing in
accordance with equation (\ref{temp}), so that the energy density
of the photon gas scales as $\rho_{\gamma} \sim a^{-4}$.

Starting at the present day, roughly 10 or 15 billion
years after the big bang, let us go back through the
history of the universe. With time reversed, we see the universe
contracting and the temperature increasing. Roughly
$t\simeq 300,000$ years from the start, the
temperature has reached several thousand degrees Kelvin. Electrons get
stripped from the atoms and the universe is filled with a hot plasma.
Further back in time, at $t\simeq 1$ second, the
temperature gets so high that the atomic nuclei
break up into their constituent protons and neutrons. Our knowledge of
nuclear physics tells us this happens at a temperature of
$10^{10}$ $^\circ$K. At this point let us stop going back
and let time move forward again.
The story resumes with the universe filled by a hot, dense
soup of neutrons, protons, and electrons. As the universe expands the
temperature drops. Within the first minute the temperature drops to
$10^{9}$ $^\circ$K and the neutrons and protons
begin to fuse together to produce the nuclei of the light elements
deuterium, helium, and lithium.
In order to produce the abundances seen
today, the nucleon density must have been roughly $10^{18}\; {\rm cm}^{-3}$.
Today we observe a nucleon density of $\sim 10^{-6}\, {\rm cm}^{-3}$, which
tells us the universe has expanded by a factor of roughly $(10^{18}/
10^{-6})^{1/3}=10^{8}$. Using equation (\ref{temp}), we therefore
expect the photon gas today to be
at a temperature of roughly $10^\circ$K. 
George Gamow made this back-of-the-envelope prediction in 1946.

In 1965, Penzias and Wilson discovered a highly uniform background of
cosmic microwave radiation at a temperature of $\sim 3^\circ$K.
This {\it cosmic microwave background} (CMB) is quite literally
the afterglow of the big bang.
More refined nucleosynthesis calculations predict a photon temperature
of $\sim 3^\circ$K, and more refined measurements of the CMB reveal it
to have a black body spectrum at a temperature of
$T_0 = 2.728\pm 0.010 $ $^\circ$K.
Typical cosmic microwave photons have wavelengths
roughly equal to the size of the letters on this page.

The CMB provides strong evidence for the
homogeneity and isotropy of space. If we look out in any direction
on the sky, we see the same microwave temperature to 1 part in $10^4$.
This implies the curvature of space is also constant to 1 part
in $10^4$ on large scales.  This observed homogeneity lets
cosmologists approximate the large-scale structure of the universe
not by a general spacetime, but by one having well defined
spatial cross sections of constant curvature.
In these Friedman-Robertson-Walker (FRW) models,
the spacetime manifold ${\mathcal M}$ is topologically the product
$\mathbb{R}\times \Sigma$ where the real line $\mathbb{R}$ represents
time and $\Sigma$ represents a 3-dimensional space of constant
curvature.\footnote{Even though elementary particle theory suggests
the universe is orientable, both the present article and the
research program of the authors and their colleagues permit
nonorientable universes as well.}
The metric on the spacelike slice $\Sigma(t)$ at time $t$ is given
by the scale factor $a(t)$ times the standard metric of constant
curvature $k =+1, 0,\, -1$.
The sectional curvature is $k/a(t)^2$, so when $\vert k \vert = 1$,
the scale factor $a(t)$ is the curvature radius;
when $k = 0$ the scale factor remains arbitrary.

The function $a(t)$ describes the evolution of the universe.
It is completely determined by Einstein's field equation.
In general Einstein's equation is a tensor equation in spacetime,
but for a homogeneous and isotropic spacetime it reduces to
the ordinary differential equation
\begin{equation}\label{fried}
\left({\dot a \over a}\right)^2+{k \over a^2}= {8 \pi G \over
3}\,  \rho \, .
\end{equation}
Here $G$ is Newton's gravitational constant,
$\rho$ is the mass-energy density, $\dot a = da/dt$,
and we have chosen units that make the speed of light $c=1$.

The first term in equation (\ref{fried}) is the
Hubble parameter $H = \dot a / a$, which tells how fast
the universe is expanding or contracting.
More precisely, it tells the fractional rate of change
of cosmic distances.  Its current value $H_0$,
called the {\it Hubble constant}, is about
65 (km/sec)/Mpc.\footnote{The abbreviation Mpc denotes a megaparsec, or
one million parsecs.  A parsec is one of those strange units invented
by astronomers to baffle the rest of us. One parsec defines the
distance from Earth of a star whose angular position shifts by
1 second of arc over a 6 month period of observation. {\it i.e.} a
parsec is defined by parallax with two Earth-sun radii
as the baseline.  A parsec is about 3 light-years.}
Thus, for example, the distance to a galaxy
100 Mpc away would be increasing at about 6500 km/sec,
while the distance to a galaxy 200 Mpc away would be increasing
at about 13000 km/sec.

Substituting $H = \dot a / a$ into equation (\ref{fried})
shows that when $k = 0$
the mass-energy density $\rho$ must be exactly $3 H^2 / 8 \pi G$.
Similarly, when $k = +1$ (resp. $k = -1$),
the mass-energy density $\rho$ must be greater than
(resp. less than) $3 H^2 / 8 \pi G$.
Thus if we can measure the current density
$\rho_0$ and the Hubble constant $H_0$ with sufficient precision,
we can deduce the sign $k$ of the curvature.  Indeed, if $k \neq 0$
we can solve for the curvature radius
\begin{equation}\label{curv}
a
= {1 \over H} \, \sqrt{ {k \over 8 \pi G \rho / 3 H^2 - 1} }
= {1 \over H} \, \sqrt{ {k \over \Omega - 1} }\, ,
\hspace*{0.5in} k\neq 0 \, ,
\end{equation}
where the {\it density parameter} $\Omega$ is the dimensionless ratio
of the actual density $\rho$ to the critical density
$\rho_c = 3 H^2 / 8 \pi G$.

The universe contains different forms of mass-energy, each of which
contributes to the total density:
\begin{equation}
\Omega = {\rho \over \rho_c} = {\rho_\gamma +\rho_m +\rho_\Lambda
\over \rho_c} = \Omega_\gamma+\Omega_m+\Omega_\Lambda \, ,
\end{equation}
where $\rho_\gamma$ is the energy density in radiation,
$\rho_m$ is the energy density in matter, and $\rho_\Lambda$
is a possible vacuum energy.
Vacuum energy appears in many current theories of the very early universe,
including the inflationary paradigm.
Vacuum energy with density $\rho_{\Lambda}= 3 \Lambda / 8\pi G$
mimics the cosmological constant $\Lambda$,
which Einstein introduced into his field equations in 1917 to avoid
predicting an expanding or contracting universe, and later retracted
as ``my greatest blunder''.

In a universe containing only ordinary matter
($\Omega_\gamma = \Omega_\Lambda = 0$),
the mass density scales as $\rho = \rho_0 (a_0 / a)^3$.
Substituting this into equation (\ref{fried}),
one may find exact solutions for $a(t)$.
These solutions predict that
if $\Omega < 1$ the universe will expand forever,
if $\Omega > 1$ the expansion will slow to a halt and the
universe will recontract, and in the borderline case $\Omega = 1$
the universe will expand forever, but at a rate $\dot a$ approaching zero.
These predictions make good intuitive sense:
under the definition of $\Omega$ as the ratio
$8 \pi G \rho / 3 H^2$, $\Omega > 1$ means the mass density $\rho$
is large and/or the expansion rate $H$ is small, so the gravitational
attraction between galaxies will slow the expansion to a halt and
bring on a recollapse;
conversely, $\Omega < 1$ means the density is small and/or the expansion
rate is large, so the galaxies will speed away from one other faster
than their ``escape velocity''.

Cosmologists have suffered from a persistent misconception
that a negatively curved universe must be the infinite hyperbolic
3-space $\mathbb{H}^3$.  This has led to the unfortunate habit of
using the term ``open universe'' to mean three different things:
``negatively curved'', ``spatially infinite'', and ``expanding forever''.
Talks have even been given on the subject of ``closed open models'',
meaning finite hyperbolic 3-manifolds
(assumed to be complete, compact, and boundaryless).
Fortunately, as finite manifolds are becoming more widely
understood, the terminology is moving towards the following.
Universes of positive, zero, or negative spatial curvature
({\it i.e.} $k = +1,0,-1$) are called ``spherical'', ``flat'',
or ``hyperbolic'', respectively.
Universes that recollapse, expand forever with zero limiting velocity,
or expand forever with positive limiting velocity are called
``closed'', ``critical'', or ``open'', respectively.
(Warning:  This conflicts with topologists' definitions
of ``closed'' and ``open''.)

\section*{Messengers from the edge of time}

In the early 1990s, the COBE satellite detected small intrinsic
variations in the cosmic microwave background temperature,
of order 1 part in $10^5$.
This small departure from perfect isotropy is thought to be due
mainly to small variations in the mass distribution of the early universe.
Thus the CMB photons provide a fossil record of the big bang.
The field of ``cosmic paleontology'' is set for a major boost
in the next decade as NASA plans to launch the
{\em Microwave Anisotropy Probe} (MAP) and ESA the
{\em Planck Surveyor}. These satellites will produce clean
all-sky maps of the microwave sky with one fifth of a degree resolution.
In contrast, the COBE satellite produced a very noisy map at
ten degree resolution \cite{bennet}. But where are the CMB photons
coming from, and what can they tell us about the curvature
and topology of space?

The first thing to realize about any observation in cosmology is that
one cannot talk about ``where'' without also talking about ``when''.
By looking out into space we are also looking back in time,
as all forms of light travel at the same finite speed.
Recall that at about 300,000 years after the big bang,
at a redshift of $z \simeq 1400$,
the entire universe was filled with an electron-ion-photon plasma
similar to the outer layers of a present-day star.
In contrast to a gas of neutral atoms, a charged
plasma is very efficient at scattering light, and is therefore opaque.
We can see back to, but not beyond, the surface of last scatter
at $z \simeq 1200$.

\begin{figure}[h]
\vspace*{60mm}
\includegraphics{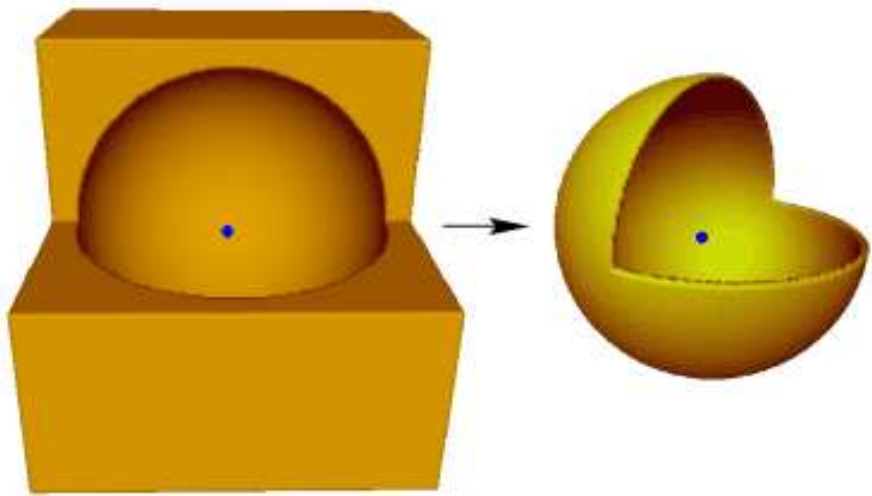}
\vspace*{3mm}
\caption{(Left) A block of space at the time of last scatter sliced
open to show the surface of last scatter seen by us today. The dot
marks the point where the Earth will eventually form. (Right) A
cut-away view showing the spherical shell we refer to as the last
scattering surface.}
\end{figure}

Once the plasma condensed to a gas, the universe became transparent
and the photons have been travelling largely unimpeded ever since.
They are distributed homogeneously throughout the universe,
and travel isotropically in all directions.
But the photons we measure with our instruments
are the ones arriving {\em here} and {\em now}.
Our position defines a preferred point in space and our
age defines how long the photons have been travelling to get here.
Since they have all been travelling at the same speed for
the same amount of time, they have all travelled the same distance.
Consequently, the CMB photons we measure today originated on a 2-sphere
of fixed radius, with us at the center (see Figure 1).
An alien living in a galaxy a billion light years away would see a
different sphere of last scatter.

To be precise, the universe took a finite amount of time to make the
transition from opaque to transparent (between a redshift of $z\simeq
1400$ and $z \simeq 1200$), so the sphere of last scatter is more
properly a spherical shell with a finite thickness. However, the
shell's thickness is less than 1\% of its radius, so cosmologists
usually talk about the sphere of last scatter rather than the shell of
last scatter.

\begin{figure}[h]
\vspace*{75mm}
\includegraphics{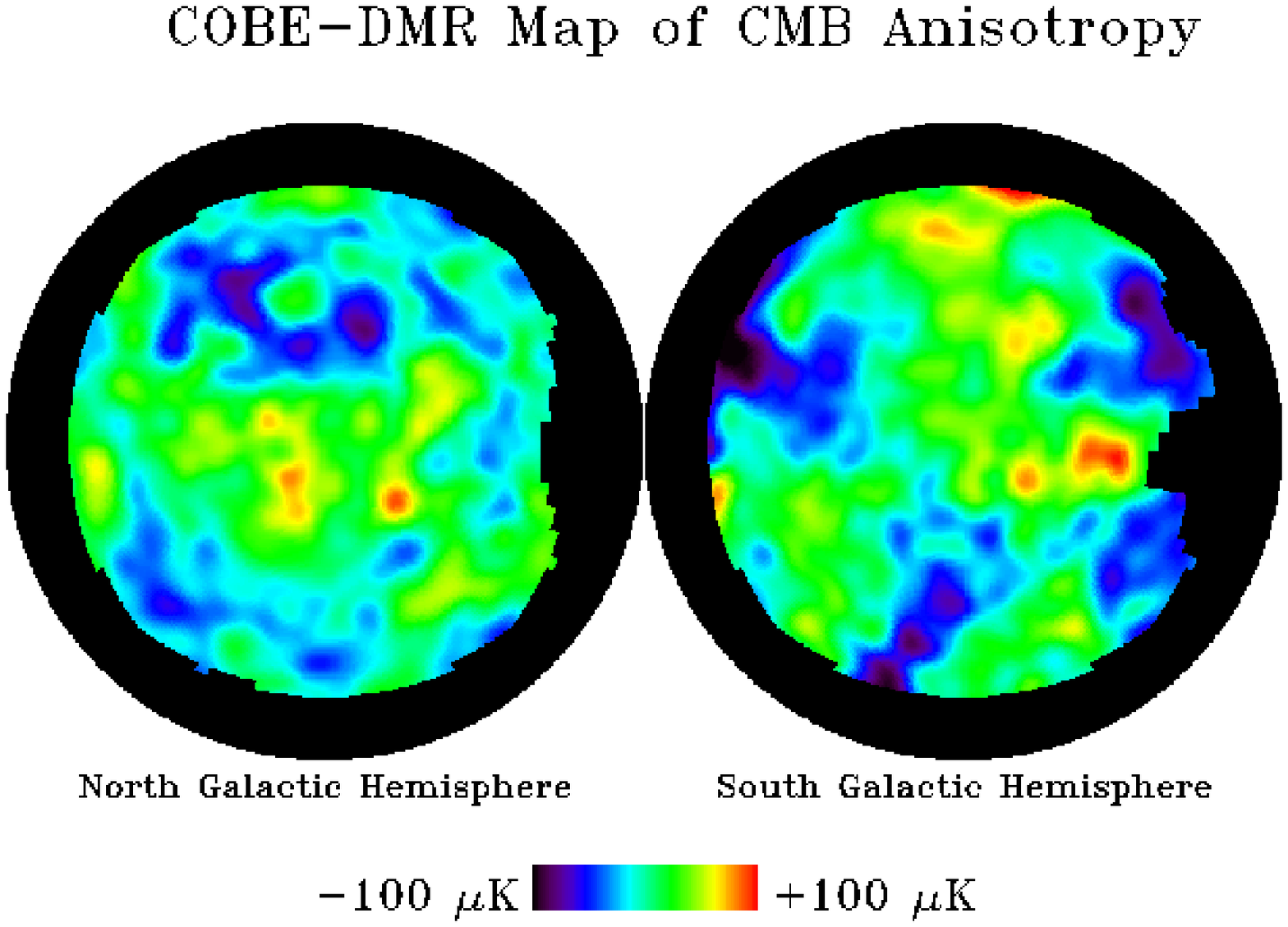}
\caption{The temperature variations in the CMB measured by the COBE
satellite. The northern and southern hemispheres of the celestial
sphere have been projected onto flat circular disks. Here the
equator is defined by the galactic plane of the Milky Way. The black
pixels correspond
to portions of the sky where the data was badly contaminated by galactic
emissions.}
\end{figure}

Now that we know where (and when) the CMB photons are coming from,
we can ask what they have to tell us. In a perfectly homogeneous and
isotropic universe, all the CMB photons would arrive here with exactly
the same energy. But this scenario is ruled out by our very existence.
A perfectly homogeneous and isotropic universe expands to
produce a perfectly homogeneous and isotropic universe. There would be
no galaxies, no stars, no planets, and no cosmologists. A more plausible
scenario is to have small inhomogeneities in the early universe
grow via gravitational collapse to produce the structures we see
today. These small perturbations in the early universe will cause
the CMB photons to have slightly different energies:  photons coming
from denser regions have to climb out of deeper potential wells, and
lose some energy while doing so. There will
also be line-of-sight effects as photons coming from
different directions travel down different paths and
experience different energy shifts. However, the energy shifts
en route are typically much smaller than the intrinsic
energy differences imprinted at last scatter. Thus, by making a map of
the microwave sky we are making a map of the density distribution on a
2-dimensional slice through the early universe. Figure 2 shows the
map produced by the COBE satellite \cite{bennet}.
In the sections that follow, we will explain how we can use such maps
to measure the curvature and topology of space. But first we need to
say a little more about geometry and topology.

\section*{Geometry and topology}
\subsection*{Topology determines geometry}

For ease of illustration, we begin with a couple of 2-dimensional
examples.

\begin{figure}[h]
\vspace*{60mm}
\includegraphics{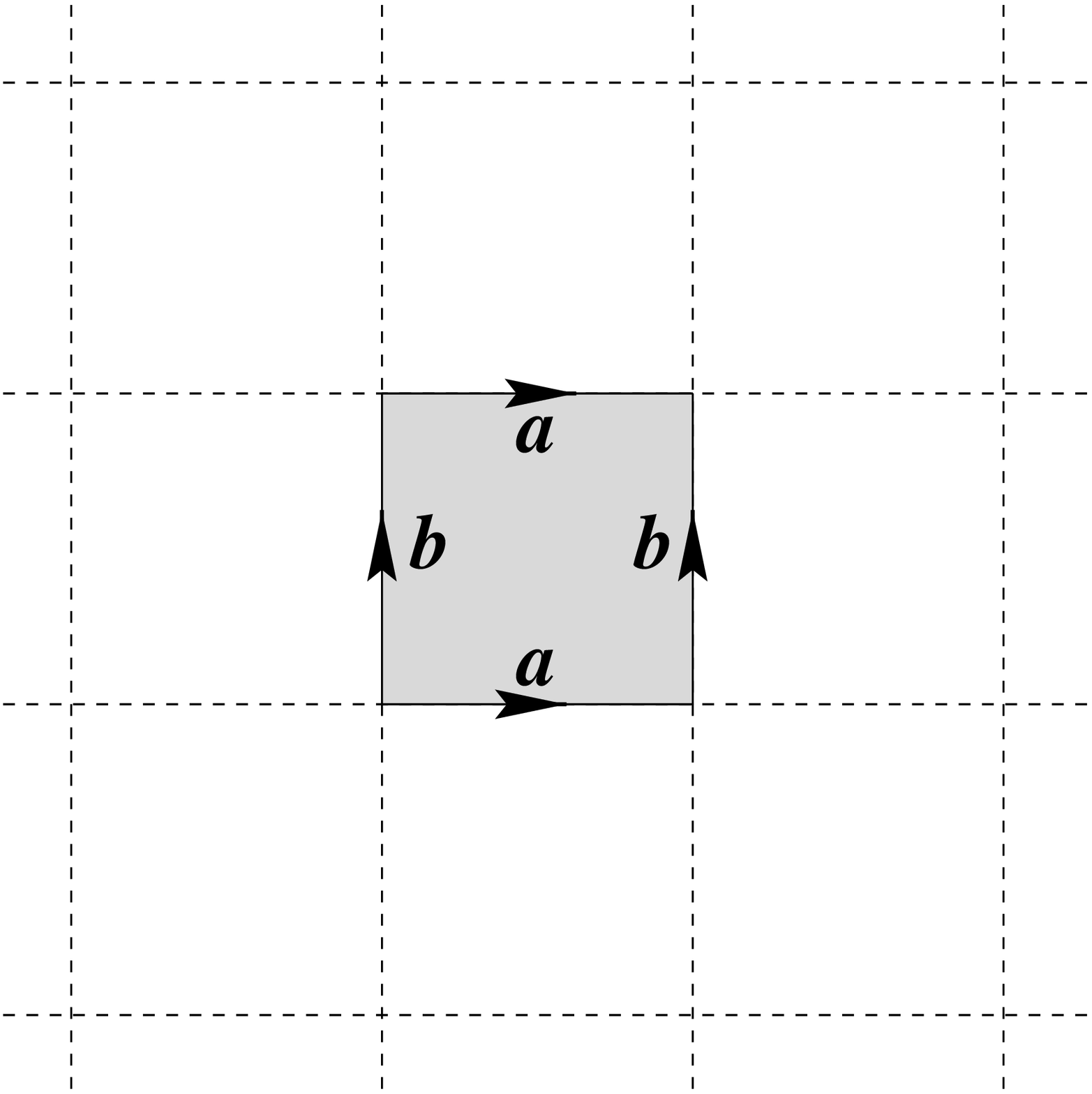}
\includegraphics{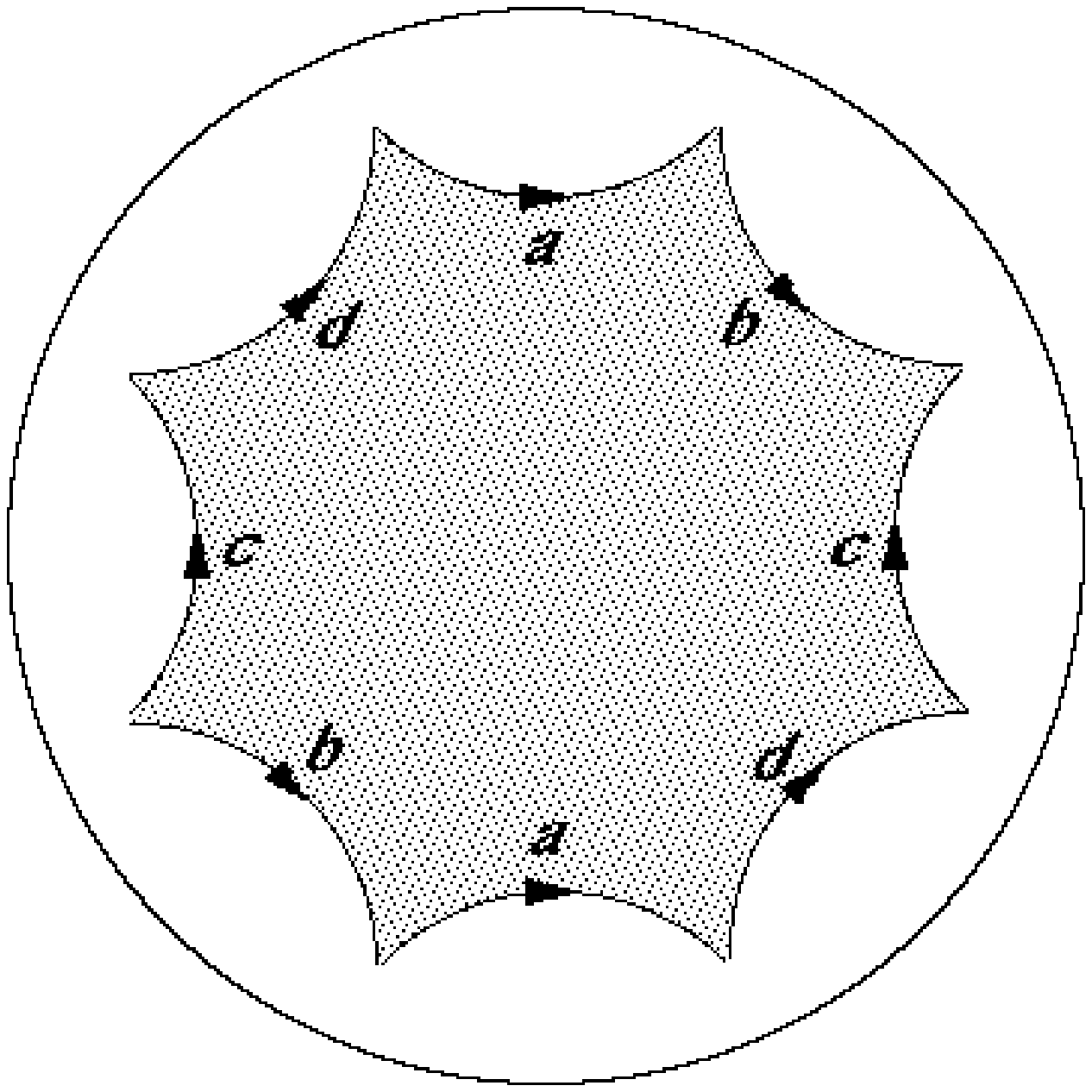}
\caption{(a) Flat torus. (b) Closed hyperbolic manifold. }
\end{figure}

A flat torus (Figure 3a) may be constructed as either a square
with opposite sides identified (the ``fundamental domain'' picture),
or as the Euclidean plane modulo the group of motions generated
by $x \rightarrow x+1$ and $y \rightarrow  y+1$
(the ``quotient picture'').  Similarly,
an orientable surface of genus two (Figure 3b) may be constructed as
either a regular hyperbolic octagon with opposite sides identified,
or as the hyperbolic plane modulo a certain discrete group of motions.
In this fashion, every closed surface may be given a geometry
of constant curvature.

Note that in the construction of the torus, the square's four
corners come together at a single point in the manifold itself,
so it is crucial that the square's angles be exactly $(2\pi)/4 = \pi/2$.
In other words, it is crucial that we start with a Euclidean square --
a hyperbolic square (with corner angles less than $\pi/2$) or
a spherical square (with corner angles greater than $\pi/2$)
would not do.  Similarly, in the construction of the genus-two surface,
the octagon's eight corners come together at a single point in the
manifold itself, so it is crucial that the angles be exactly
$(2\pi)/8 = \pi/4$.  A Euclidean or spherical octagon would not do.
In fact, even a smaller (resp. larger) hyperbolic octagon would
not do, because its angles would be greater (resp. less) than $\pi/4$.
More generally, in any constant curvature surface,
the Gauss-Bonnet theorem $\int k \, \vert dA \vert = 2\pi\chi$ forces the sign
of the curvature $k$ to match the sign of the Euler number $\chi$.

The constructions of Figure 3 generalize to three dimensions.
For example, a 3-torus may be constructed as either a cube with opposite
faces identified, or as the 3-dimensional Euclidean space $\mathbb{E}^3$ modulo
the group generated by $x \rightarrow x+1$, $y \rightarrow  y+1$,
$z \rightarrow z+1$.
Similar constructions yield hyperbolic and spherical manifolds.
In the spherical and hyperbolic cases, the connection between the
geometry and the topology is even tighter than in two dimensions.
For spherical and hyperbolic 3-manifolds, the topology {\em completely}
determines the geometry, in the sense that if two spherical or
hyperbolic manifolds are topologically equivalent (homeomorphic),
they must be geometrically identical (isometric) as
well.\footnote{In the hyperbolic case this result is a special case
of the Mostow Rigidity Theorem.}
In other words, spherical and hyperbolic 3-manifolds are rigid.
However, this rigidity does not extend to Euclidean 3-manifolds:
a 3-torus made from a cube and a 3-torus made from a parallelepiped
are topologically equivalent, but geometrically distinct.
The final section of this article will explain how the rigidity
of a closed spherical or hyperbolic universe may be used to refine
the measured radius of the last scattering sphere.

When we look out into the night sky, we may be seeing multiple
images of the same finite set of galaxies, as Figure 3
suggests.  For this reason cosmologists studying
finite universes make heavy use of the ``quotient picture'' described
above, modelling a finite universe as hyperbolic 3-space, Euclidean
3-space, or the 3-sphere, modulo a group of rigid motions.
If we could somehow determine the position and orientation
of all images of, say, our own galaxy, then we would know the group of
rigid motions, and thus the topology.  Unfortunately we cannot
recognize images of our own galaxy directly.  If we are seeing it at
all, we are seeing it at different times in its history, viewed
from different angles -- and we do not even know what it looks like
from the outside in any case.  Fortunately we {\em can} locate
the images of our own galaxy indirectly, using the cosmic microwave
background. The section {\it Observing the topology of the universe}
will explain how.

\subsection*{Geometric models}

Elementary linear algebra provides a consistent way to model
the 3-sphere, hyperbolic 3-space, and Euclidean 3-space.

Our model of the 3-sphere $\mathbb{S}^3$ is the standard one.
Define Euclidean 4-space $\mathbb{E}^4$ to be the vector space
$\mathbb{R}^4$ with the usual inner product
$\langle u,v \rangle = u_0 v_0 + u_1 v_1 + u_2 v_2 + u_3 v_3$.
The 3-sphere is the set of points one unit from the origin,
{\it i.e.} $\mathbb{S}^3 = \{v \mid \langle v,v \rangle = 1\}$.
The standard $4 \times 4$ rotation and reflection matrices
studied in linear algebra naturally represent
rigid motions of $\mathbb{S}^3$, both in theoretical discussions
and in computer calculations. These matrices generate the
orthogonal group $O(4)$.

Our model of the hyperbolic 3-space $\mathbb{H}^3$ is formally almost
identical to our model of the 3-sphere.  Define Minkowski space
$\mathbb{E}^{1,3}$ to be the vector space $\mathbb{R}^4$ with the inner product
$\langle u,v \rangle = -u_0 v_0 + u_1 v_1 + u_2 v_2 + u_3 v_3$
(note the minus sign!).
The set of points whose squared ``distance'' from the origin is $-1$
is, to our Euclidean eyes, a hyperboloid of two sheets.
Relative to the Minkowski space metric, though, each sheet is a copy
of hyperbolic 3-space.
Thus our formal definition is
$\mathbb{H}^3 = \{v \mid \langle v,v \rangle = -1, v_0 > 0\}$.
The rigid motions of $\mathbb{H}^3$ are represented by the
``orthogonal matrices'' that preserve both the Minkowski space
inner product and the sheets of the hyperboloid.
They comprise an index 2 subgroup of the Lorentz group $O(1,3)$.

The tight correspondence between our models for $\mathbb{S}^3$
and $\mathbb{H}^3$ extends only partially to the Euclidean
3-space $\mathbb{E}^3$.  Borrowing a technique from the computer
graphics community,
we model $\mathbb{E}^3$ as the hyperplane at height 1
in $\mathbb{E}^4$, and represent its isometries as the subgroup
of $GL_4(\mathbb R)$ that takes the hyperplane rigidly to itself.

This formal correspondence between $\mathbb{S}^3$ and $\mathbb{H}^3$ reveals
spherical and hyperbolic geometry to be surprisingly similar.
Any theorem you prove about one (using only linear algebra
for the proof) translates to a corresponding theorem about
the other.  Not only does the statement of the theorem transfer
from one geometry to the other, but the proof itself may be
copied line by line, inserting or removing minus signs as necessary.
For example, the proof of the spherical Law of Cosines
\begin{equation}\label{scos}
\cos C = \cos A \, \cos B + \sin A \, \sin B\, \cos \gamma
\end{equation}
translates mechanically to a proof for the hyperbolic Law of Cosines
\begin{equation}\label{hcos}
\cosh C = \cosh A \, \cosh B - \sinh A \, \sinh B\, \cos \gamma
\end{equation}
where $A$, $B$ and $C$ are the lengths of a triangle's sides,
and $\gamma$ is the angle opposite side $C$.
{\it N.B.} The spherical and hyperbolic trig functions share
a direct correspondence.  The functions $\cos d$ and $\sin d$ are,
by definition, the coordinates of the point one arrives at after
travelling a distance $d$ along the unit circle $\mathbb{S}^1$.
It is true that $\cos d = (e^{id} + e^{-id})/2$ and
$\sin d = (e^{id} - e^{-id})/(2i)$,
but that is most naturally a theorem, not a definition.
Similarly, the functions $\cosh d$ and $\sinh d$ are, by definition,
the coordinates of the point one arrives at by
travelling a distance $d$ along the unit hyperbolic $\mathbb{H}^1$
(measure the distance with the native Minkowski space metric,
not the Euclidean one!).
It is true that $\cosh d = (e^d + e^{-d})/2$ and
$\sinh d = (e^d - e^{-d})/2$,
but that is most naturally a theorem, not a definition.

Euclidean geometry does not correspond nearly so tightly to
spherical and hyperbolic geometry as the latter two do to each other.
Fortunately the theorems of Euclidean geometry can often be obtained
as limiting cases of the theorems of spherical or hyperbolic geometry,
as the size of the figures under consideration approaches zero.
For example, the familiar Euclidean Law of Cosines
\begin{equation}
C^2 = A^2 + B^2 - 2AB \cos \gamma
\end{equation}
may be obtained from (\ref{scos}) by substituting the small angle
approximations $\cos x \approx 1 - x^2/2$ and $\sin x \approx x$,
or from (\ref{hcos}) by substituting $\cosh x \approx 1 + x^2/2$ and
$\sinh x \approx  x$.

\subsection*{Natural units}

Spherical geometry has a natural unit of length.  Commonly
called a radian, it is defined as the 3-sphere's radius in the
ambient Euclidean 4-space $\mathbb{E}^4$.  Similarly, hyperbolic geometry
also has a natural unit of length.  It too should be called
a radian, because it is defined as (the absolute value of) the
hyperbolic space's radius in the ambient Minkowski space $\mathbb{E}^{1,3}$.
The scale factor $a(t)$ introduced in the {\it Big bang cosmology}
section connects the mathematics to the physics: it tells,
at each time $t$, how many meters correspond to one radian.
In both spherical and hyperbolic geometry the radian is more
often called the {\em curvature radius}, and quantities reported
relative to it are said to be in {\em curvature units}.
Euclidean geometry has no natural length scale, so all measurements
must be reported relative to some arbitrary unit.

\section*{Measuring the curvature of the universe}

To determine the curvature of the universe,
cosmologists seek accurate values for the
Hubble constant $H_0$ and the density parameter $\Omega_0$.
If $\Omega_0 = 1$, space is flat.
Otherwise equation (\ref{curv}) tells the curvature radius $a_0$.
The parameters $H_0$ and $\Omega_0$ may be deduced from
observations of ``standard candles'' or the CMB.

Before describing these cosmological measurements, it is helpful
to consider how one might measure the curvature of space in a
static universe.

\subsection*{Ideal measurements in a static universe}

Gauss taught us that we can measure the curvature of a surface by
making measurements {\em on} the surface. He showed that the sum of the
interior angles in a triangle depended on two factors - the curvature
of the surface and the size of the triangle. On a flat surface the
angle sum is $180^o$, on a positively curved surface the sum is
greater than $180^o$, and on a negatively curved surface the angle sum
is less than $180^o$. The deviation from the Euclidean result depends
on the size of the triangle relative to the radius of curvature. The
larger the triangle the larger the deviation. Another way to
characterise the curvature of a surface is to look at how the
circumference of a circle varies with radius. For a
flat surface a circle's circumference grows linearly with radius.
On a positively
curved surface the circumference grows more slowly, and on a
negatively curved surface the circumference grows more quickly.

\begin{figure}[h]
\vspace*{65mm}
\includegraphics{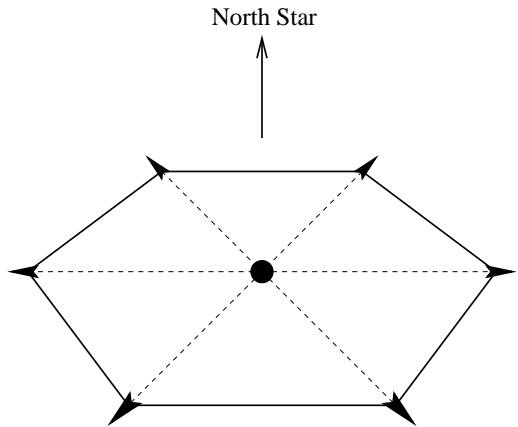}
\caption{Determining the curvature by measuring the distance between
six equally spaced rockets and comparing it to their distance from earth.}
\end{figure}

These ideas can easily be generalized to higher dimensions. In the
delightful book {\it Poetry of the
Universe}, Bob Osserman\cite{bob} describes a direct
method for measuring the curvature of space. He envisions launching
six equally spaced rockets in a plane around the equator, each travelling
at the same speed away from the earth (see Figure 4). In a flat
universe the distance between each pair of rockets will remain equal
to their distance from the earth. In a
positively curved space the distance between the rockets would grow
more slowly than their distance from the earth, while in a negatively
curved space it would grow more quickly. Osserman's method neatly
combines features of the triangle angle-sum and circle circumference
methods. The rockets lie on a circle, so the growth in distance
between them reflects the growth in the circle circumference. In
addition, the angle between the lines connecting each pair of
rockets will be greater than or less than $2\pi /3$ (the flat space
result) if space has positive or negative curvature respectively.

Unfortunately, the sheer size of the universe makes this elegant test
unworkable. Existing observations tell us that the curvature radius is
at least 3000 Mpc, so it would take rockets travelling near the speed
of light billions of years to form triangles large enough to reveal
the curvature. However, very similar measurements of the curvature can
be made using objects that are already billions of light years away.

\subsection*{Standard candle approach}

Astronomers observe objects like Cepheid variables and type Ia supernovae
whose intrinsic luminosity is known (many question this assertion).
In a static Euclidean space, the apparent brightness of such standard
candles would fall off as the square of their distance from us.
The reduction in brightness occurs as the photons spread out over
a surface who's area grows as $4 \pi$ times the square of the distance
travelled. In a static spherical space the apparent brightness would fall-off
more slowly as the area grows more slowly; in a static hyperbolic
space the apparent brightness would fall-off more quickly as the area
grows more quickly. This is the higher dimensional analog of the
circumference-radius method used to measure the curvature of a surface.

In an expanding universe the distance-brightness relationship is more
complicated. Different values of $H_0$ and $\Omega_0$ predict different
relationships between a standard candle's apparent brightness ${\mathcal F}$
and its redshift $z$.
Sufficiently good observations of ${\mathcal F}$ and $z$
for sufficiently good standard candles will tell us the values
of $H_0$ and $\Omega_0$, and thence the curvature radius $a_0$.

The results of standard candle observations are still inconclusive.
Some recent measurements \cite{perl} have yielded results inconsistent
with the assumption of a matter-dominated universe, and point instead
to a vacuum energy $\Omega_\Lambda$ exceeding the density $\Omega_m$
of ordinary matter! More refined measurements over the next few years
and a better understanding of the standard candles should settle this issue.

\subsection*{CMB approach}

The temperature fluctuations in the CMB (recall Figure 2) are,
to a mathematician, a real-valued function on a 2-sphere.
As such they may be decomposed into an infinite series
of spherical harmonics, just as a real-valued function on a circle
may be decomposed into an infinite series of sines and cosines.
And just as the Fourier coefficients of a sound wave provide much
useful information about the sound (enabling us to recognize it as,
say, the note A$\flat$ played on a flute), the Fourier coefficients
of the CMB provide much useful information about the dynamics of
the universe.  In particular, they reflect the values of
$H_0$, $\Omega_o$, $\Omega_\Lambda$, and other cosmological
parameters.

It is interesting to look at how the ``sounds of the CMB'' can be used
to measure the total energy density $\Omega_o$. The method uses Gauss'
triangles on the largest possible scale. We expect
to find a peak in the CMB anisotropy corresponding to the angle
subtended by the Hubble radius, $R_H=H^{-1} = a /\dot a$, at the time
of last scatter. By measuring this angle and using using the Law of
Cosines, we can fix $\Omega_o$.

Taking into account the expansion of the universe,
the Hubble radius at last scatter corresponds to a distance
of $L_H \simeq H_o^{-1}/\sqrt{ \Omega_o \, z_{sls}}$ (roughly
$100 \rightarrow
200$ Mpc today). Within the confines of a Hubble patch, waves are free
to propagate in the electron-ion plasma. Outside a Hubble patch
space is expanding faster than the speed of light and causality
prevents the plasma from oscillating.
Waves inside the Hubble patch give the CMB photons a kick that
either increases or decreases their energy. Thus, on scales smaller
than $L_H$ we expect to see an increased anisotropy in the CMB sky.

\begin{figure}[h]
\vspace*{70mm}
\includegraphics{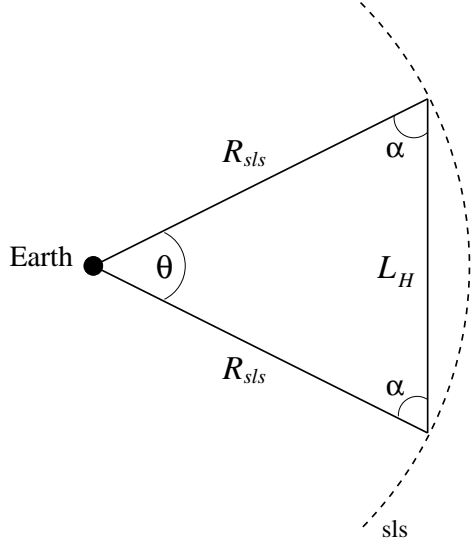}
\caption{The cosmic triangle used to measure the curvature.}
\end{figure}

Using the MAP satellite we will be able to measure the angular scale
$\theta$ subtended by the length $L_H$, at a distance equal to
the radius of the surface of last scatter, $R_{sls}$ (see Figure 5).
The distance $R_{sls}$ can be found by following a photon's trajectory
in a FRW spacetime between last scatter and today. In a matter
dominated universe the result is
\begin{equation}\label{sls}
R_{sls} \simeq a_o\, {\rm arcsinh}\left({2 \sqrt{1-\Omega_o} \over
\Omega_o}\right) \, .
\end{equation}
We now have all the ingredients in place to measure the angle sum in
the cosmic triangle. Notice that Einstein's equation has been used
to recast the side lengths in terms of the matter density $\Omega_o$.
Scaling these lengths by the curvature radius $a_o$, we can use either
the spherical or hyperbolic Law of Cosines to relate the angular size
of the Hubble patch to the density parameter:
\begin{equation}
\theta \approx \sqrt{{ \Omega_o \over   4 z_{sls}}}\, \quad \Rightarrow
\quad \theta \approx 0.83^o
\times \sqrt{\Omega_o} \, .
\end{equation}
Here we have used a small angle approximation and worked
to leading order in $z_{sls}^{-1}$. As expected,
the angle $\theta$ will be smaller in a hyperbolic universe $\Omega_o <1$
and larger in a spherical universe $\Omega_o>1$. 
It is amusing to go a step further and use the Law of Sines
to solve for the other interior angle, $\alpha$, and from
this find the interior angle sum $\Sigma = \theta+2 \alpha$
in terms of the density parameter:
\begin{equation}
\Sigma = \pi - { 1 - \Omega_o \over \sqrt{\Omega_o  \,
 z_{sls} }} \, .
\end{equation}
We see that the sum of the interior angles is less than $180^o$ in
hyperbolic space and greater than $180^o$ in spherical space.
Current astronomical observations suggest $\Omega_o \approx 0.3$. If
these observations are true, the MAP satellite
will discover an angular deficit of $2.2^o$ for our cosmic triangle!

\section*{Observing the topology of the universe}

In a finite universe we may be seeing the same set of
galaxies repeated over and over again. Like a hall of mirrors, a
finite universe gives the illusion of being infinite. The illusion
would be shattered if we could identify repeated images of some
easily recognizable object. The difficulty is finding objects that
can be recognized at different times and at different orientations.
A more promising approach is to look for correlations in the cosmic
microwave background radiation. The CMB photons all originate from the
same epoch in the early universe, so there are no aging effects to
worry about. Moreover, the shell they originate from is very thin, so
the surface of last scatter looks the same from either side. The
importance of this second point will soon become clear.

\begin{figure}[h]
\vspace*{75mm}
\includegraphics{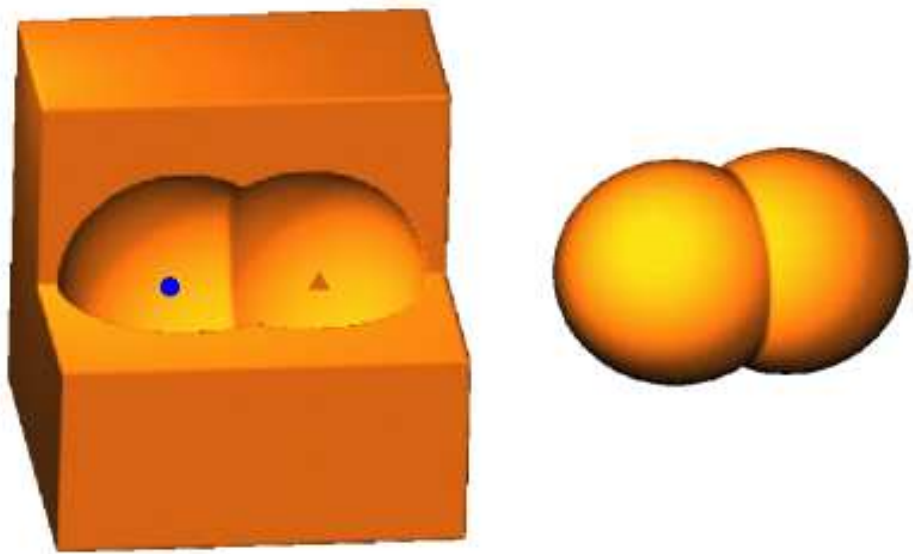}
\vspace*{3mm}
\caption{(Left) A block of space at the time of last scatter sliced
open to show two different surfaces of last scatter. The dot
marks our vantage point and the triangle marks the alien's vantage
point. (Right) An outside view showing how our shell of last scatter
intersects the alien's.}
\end{figure}

Consider for a moment two different views of the universe: one from
here on Earth and the other from a faraway galaxy. As mentioned
earlier, an alien living in that faraway galaxy would see a
different surface of last scatter (see Figure 6). The alien's
CMB map would have a different pattern of hot and cold spots from
ours. However, so long as the alien is not too far away, our two maps
will agree along the circle defined by the intersection of our last
scattering spheres. Around this circle we would both see
exactly the same temperature pattern as the photons came from exactly
the same place in the early universe. Unless we get to exchange notes
with the alien civilization, this correlation along the matched circles
plays no role in cosmology.

But what if that faraway galaxy is just another image of the Milky
Way, and what if the alien is us?
(Recall the flat torus of Figure 3a, whose inhabitants would
have the illusion of living in an infinite Euclidean plane
containing an infinite lattice of images of each object.)
Cornish, Spergel and Starkman\cite{css} realised that in a finite
universe the matched circles can transform our view of cosmology.
For then the circles become correlations on a
single copy of the surface of last scatter, {\it i.e} the matched
circles must appear at two different locations on the CMB sky.
For example, in a universe with 3-torus topology we would see matched
circles in opposite directions on the sky. More generally the
pattern of matched circles varies according to the topology.
The angular diameter of each circle pair is fixed
by the distance between the two images. Images that are displaced
from us by more than twice the radius of the last scattering
sphere will not produce matched circles. By searching
for matched circle pairs in the CMB we may find proof that
the universe is finite.

\begin{figure}[h]
\vspace*{70mm}
\includegraphics{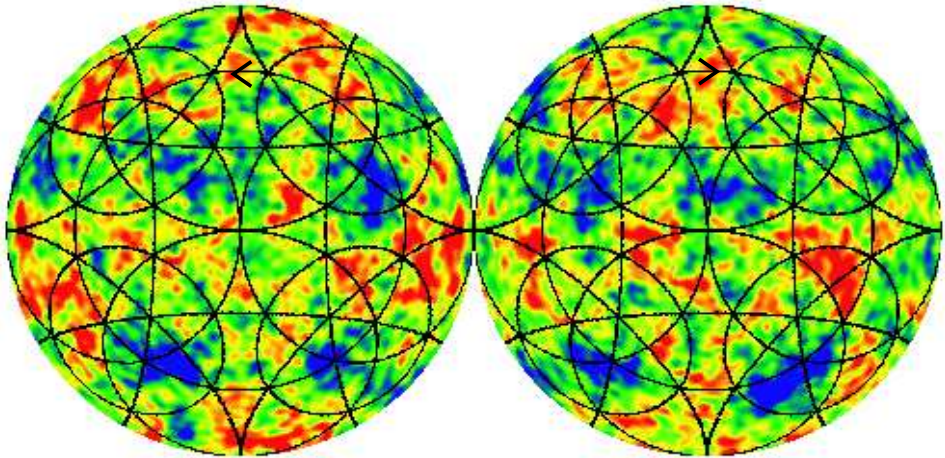}
\vspace*{3mm}
\caption{The northern and southern hemispheres of the CMB sky in a
3-torus universe. The 13 matched circle pairs are marked by black
lines. The arrows indicate the start point and relative phasing for
one pair of matched circles.}
\end{figure}

At present we do not have a good enough CMB map to perform the
search, but this will soon change. The CMB map produced by
COBE (reproduced here in Figure 2) has
a resolution of 10 degrees and 30\% of what one sees is
noise, not signal. However, by 2002 the MAP satellite will
have furnished us with a far superior map at better than $0.5^\circ$
resolution. In the interim we can test our search algorithms on
computer generated sky maps. One example of a synthetic sky map is
shown in Figure 7. The model has a cubical 3-torus topology and
a scale invariant spectrum of density perturbations. The nearest
images are separated by a distance equal to the radius of the last
scattering surface. Consequently there are 13 matched circle pairs
and 3 matched points (circles with angular diameter 0). The matched
circles are indicated by black lines.
With good eyes and a little patience one can follow the temperature
pattern around each pair of matched circles and convince oneself that
the temperatures at corresponding points are correlated.

An automated search algorithm has been developed \cite{css1} to
search for matched circle pairs.
The computer searches over all possible positions,
diameters, and relative phases. On a modern supercomputer the search
takes several hours at $1^\circ$ resolution. Our prospects for finding
matched circles are greatly enhanced if the universe is highly curved.
The rigidity described in the previous section means that the distance
between images is fixed by the curvature scale and the discrete group
of motions. We will be most interested in hyperbolic models since
observations suggest $\Omega_0\approx 0.3$.
In most low-volume hyperbolic models our
nearest images are less than one radian away. The crucial
quantity then becomes the radius of the last scattering
surface expressed in curvature units:
\begin{equation}
\eta= {R_{sls} \over a_0}
\simeq {\rm arcsinh}\left({2 \sqrt{1-\Omega_0} \over
\Omega_0}\right) \, .
\end{equation}
In a universe with $\Omega_0=0.3$ we find $\eta \approx 2.42$,
so all images less than $2\eta \approx 4.84$ radians away will produce
matched circle pairs.
However, we may have difficulty reliably detecting matched circles
with angular diameters below $\theta = 10^\circ$;
we therefore restrict ourselves to
images within a ball of radius $4.6$ radians.
In the universal cover $\mathbb{H}^3$ a sphere
of radius 4.6 encloses a volume of about 15000, so if the
universe is a hyperbolic manifold of volume less than about 100,
there will be an abundance of matched circles.

\section*{Reconstructing the topology of the universe}

If at least a few pairs of matching circles are found, they will
implicitly determine the global topology of the universe \cite{jeff}.
This section explains how to convert the list of circle pairs to
an explicit description of the topology, both as a fundamental
domain and as a quotient ({\it cf.} the section {\it Geometry
and topology}).  The fundamental
domain picture is more convenient for computing the manifold's
invariants (such as its volume, homology, etc.) and comparing
it to known manifolds, while the quotient picture is more convenient
for verifying and refining the astronomical observations.
Assume for now that space is finite, and that all circles have
been observed with perfect accuracy.

The fundamental domain we construct is a special type
known as a {\it Dirichlet domain}.  Imagine inflating a huge spherical
balloon whose center is fixed on our galaxy, and whose radius
steadily increases.  Eventually the balloon will wrap
around the universe and meet itself.  When it does,
let it keep inflating, pressing against itself just as a real
balloon would, forming a planar boundary.
When the balloon has filled the entire universe, it will have
the form of a polyhedron.  The polyhedron's faces will be
identified in pairs to give the original manifold.

Constructing a Dirichlet domain for the universe, starting
from the list of circle pairs, is quite easy.  Figure 8 shows
that each face of the Dirichlet domain lies exactly half way
between its center (our galaxy) and some other image of its center.
The previous section showed that each circle-in-the-sky also lies
exactly half way between the center of the SLS (our galaxy) and
some other image of that center. Thus, roughly speaking,
the planes of the circles and the planes of the Dirichlet domain's
faces coincide!  We may construct the Dirichlet domain as the
intersection of the corresponding half spaces.\footnote{
All but the largest circles determine planes lying wholly outside
the Dirichlet domain, which are superfluous in the intersection
of half spaces.  Conversely, if some face of the Dirichlet domain
lies wholly outside the SLS, its ``corresponding circle'' will not
exist, and we must infer the face's location indirectly.
The proof that the Dirichlet domain correctly models the topology
of the universe is, of course, simplest in the case that all
the Dirichlet domain's faces are obtained directly from observed
circles.}

\begin{figure}[h]
\vspace*{70mm}
\includegraphics{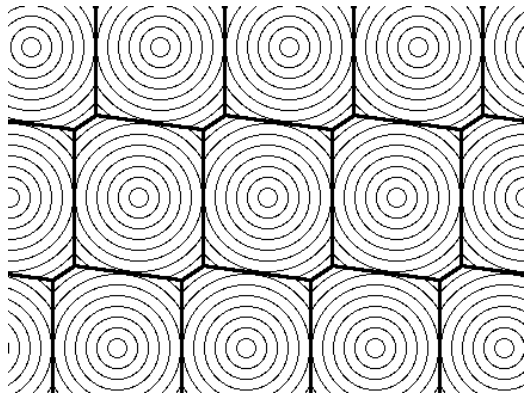}
\caption{To construct a Dirichlet domain,
       inflate a balloon until it fills the universe.}
\end{figure}

Finding the rigid motions (corresponding to the quotient picture
in the {\it Geometry and topology} section) is also easy.
The MAP satellite data will determine
the geometry of space (spherical, Euclidean or hyperbolic) and the
radius for the SLS, as well as the list of matched circles.
If space is spherical or hyperbolic, the radius of the SLS will
be given in radians ({\it cf.} the subsection {\it Natural units});
if space is Euclidean the radius will be normalized to 1.
In each case, the map from a circle
to its mate defines a rigid motion of the space, and it is
straightforward to work out the corresponding matrix
in $O(4)$, $O(1,3)$, or $GL_4(\mathbb{R})$ (recall the subsection
{\it Geometric models}).

Why bother with the matrices? Most importantly, they can verify
that the underlying data are valid. How do we know that the MAP
satellite measured the CMB photons accurately? How do we know
that our data analysis software does not contain bugs?
If the matrices form a discrete group, then we may be confident
that all steps in the process have been carried out correctly,
because the probability that bad data would define a discrete group
(with more than one generator) is zero.
In practical terms, the group is discrete if the product of any
two matrices in the set is either another matrix in the set
(to within known error bounds) or an element that is ``too far away''
to yield a circle.  More spectacularly, the matrices corresponding
to the dozen or so largest circles should predict the rest of
the data set (modulo a small number of errors), giving us
complete confidence in its validity.

We may take this reasoning a step further, and use the matrices
to correct errors.  Missing matrices may be deduced as
products of existing ones. Conversely, false matrices may
readily be recognized as such, because they will not fit into
the structure of the discrete group; that is, multiplying a
false matrix by almost any other matrix in the set will yield a
product not in the set. This approach is analogous to surveying
an apple orchard planted as a hexagonal lattice.  Even if large
portions of the orchard are inaccessible (perhaps they are
overgrown with vines), the locations of the hidden trees may be
deduced by extending the hexagonal pattern of the observable
ones. Conversely, if a few extra trees have grown between the
rows of the lattice, they may be rejected for not fitting into
the prevailing hexagonal pattern. Note that this approach will
tolerate a large number of inaccessible trees, just so the
number of extra trees is small.  This corresponds to the
types of errors we expect in the matrices describing the real
universe: the number of missing matrices will be large because
microwave sources within the Milky Way overwhelm the CMB in a
neighbourhood of the galactic equator, but the number of extra
matrices will be small (the parameters in the circle matching
algorithm are set so that the expected number of false matches
is 1).  In practice, the Dirichlet domain will not be computed
directly from the circles, as suggested above, but from the
matrices, to take advantage of the error correction.

Like all astronomical observations, the measured radius
$R_{sls}$ of the sphere of last scatter will have some error.
Fortunately, if space is spherical or hyperbolic, we can
use the rigidity of the geometry to remove most of it!
Recall that the hyperbolic octagon in Figure 3b had to be
just the right size for its angles to sum to $2\pi$.
The Dirichlet domain for the universe (determined by the
circle pairs -- {\it cf.} above) must also be just the right
size for the solid angles at its vertices to sum to
a multiple of $4\pi$.  More precisely, the face pairings
bring the vertices together in groups, and the solid
angles in each group must sum to exactly $4\pi$.
If the measured solid angle sums are consistently less than
(resp. greater than) $4\pi$, then we know that the true value
of $R_{sls}$ must be slightly less than (resp. greater than)
the measured value, and we revise it accordingly.
The refined value of $R_{sls}$ lets us refine $\Omega_0$ as well,
because the two variables depend on one another.

\section*{Acknowledgments}

Our method for detecting the topology of the universe was developed in
collaboration with David Spergel and Glenn Starkman. We are indebted
to Proty Wu for writing the visualization software used to produce
Figure 5, and to the editor for his generous help in improving
the exposition.  One of us (Weeks) thanks the U.S. National Science
Foundation for its support under grant DMS-9626780.

\end{document}